\documentclass[a4paper]{article}

\usepackage[english]{babel}
\usepackage[utf8x]{inputenc}
\usepackage{svg}
\usepackage[T1]{fontenc}

\usepackage[a4paper,top=3cm,bottom=2cm,left=3cm,right=3cm,marginparwidth=1.75cm]{geometry}

\usepackage{amsmath}
\usepackage{graphicx}
\usepackage[colorinlistoftodos]{todonotes}
\usepackage[colorlinks=true, allcolors=blue]{hyperref}

\title{Predicting intubation support requirement of patients using Chest X-ray with Deep Representation Learning}
\author{
  Maurya, Aniket\\
  \texttt{theaniketmaurya@gmail.com}
  
}

\begin{document}
\maketitle

\begin{abstract}
Recent developments in medical imaging with Deep Learning presents an evidence of automated diagnosis and prognosis. It can also be a complement to currently available diagnosis methods.
Deep Learning can be leveraged for diagnosis, severity prediction, intubation support prediction and many similar tasks.
We present prediction of intubation support requirement for patients from the Chest X-ray using Deep representation learning. We release our source code publicly on \href{https://github.com/aniketmaurya/covid-research}{https://github.com/aniketmaurya/covid-research}.

\end{abstract}

\section{Introduction}

According to \cite{who-staff-short} the world is to see a short of 12.9 million health-care workers by 2035.
This problem of shortage of medical experts and medical equipment is more visible in this COVID-19 pandemic \cite{india-shortage} \cite{supply-short}.
COVID-19 test result time has not improved much to help slow the spread of the virus \cite{slow_test}.
There is an urgent need to upgrade the current healthcare system with recent technology advancements \cite{who-push}.

Approximately 3.2\% of COVID-19 patients required intubation and invasive ventilation support in some point of their illness \cite{vent-count}.
Sometimes COVID-19 patients crash suddenly and require immediate care for intubation and ventilation \cite{strickland}.
Providing intubation can put infection risk to the provider and every person in the room.
Some delay may occur during intubation process due to heightened anxiety and rush, this can increase the infection risk.

In this work, we present prediction of intubation support requirement to patients using Deep Representation Learning. We show that how we can leverage deep learning for taking proactive measures like arranging beds, ventilators, oxygen cylinders. It can give enough time to healthcare provider to be prepared with personal protective equipment properly.
We produce Chest X-ray representation of patients and train our learning algorithm to classify whether the person will need intubation support during the duration of his illness.

\section{Related Work and Background}

Recently a lot of work has been started in automating the disease prognosis and diagnosis task with Deep Learning. It is mainly due to recent advancements in Deep learning. Convolutional Neural Networks \cite{lenet} are now able to reach human level performance in classification and detection tasks \cite{humanlevelcnn}. \cite{rajpurkar2017chexnet} \cite{chexpert} \cite{lu2020multiobjective} detects chest pneumonia with radiologist level accuracy. Image recognition with deep learning is being used for detecting different type of diseases like diabetic retinopathy and skin cancer \cite{retinopathy}.

\subsection{Radiologist level Pneumonia Detection using Chest X-ray images}

ChexNet \cite{rajpurkar2017chexnet} can predict Pneumonia from CXRs with better F1 score than a group of experienced radiologists. It is a 121-layered deep convolutional neural network trained on ChestX-ray14 dataset \cite{Wang_2017} for classifying Chest X-ray image as Pneumatic. The dataset is comprised of 112,120 frontal-view Chest X-ray images of 30,805 unique patients. To train this network, images are first resized to 224x224 dimension and normalized by the ImageNet \cite{russakovsky2015imagenet} mean and standard deviation.
This sets a benchmark for other similar lungs disease diagnosis task. The network is able to achieve state of the art score on total 14 different lungs disease diagnosis task. It achieves F1 score of 0.435 which is higher than achieved by a group of radiologists that is 0.385.

\subsection{COVID-19 Severity prediction}

\cite{cohen2020predicting} predicts severity of COVID-19 from the chest x-ray of patients. To train this network, a representation of 1024 dimensional vector is produced from non-covid CXR pre-trained network, 18 outputs (from classification layer), 4 outputs as subset of 18 outputs (lung opacity, pneumonia, infiltration, and consolidation) and Lung opacity output is used for creating final data representation. It further evaluates the input data features for overfitting \& bias and analyses the model with Saliency map. Knowing severity can help COVID-19 patients and hospitals so that they can arrange themselves.

\section{Dataset}

\begin{figure}[h]
\centering
\includegraphics[width=0.6\textwidth]{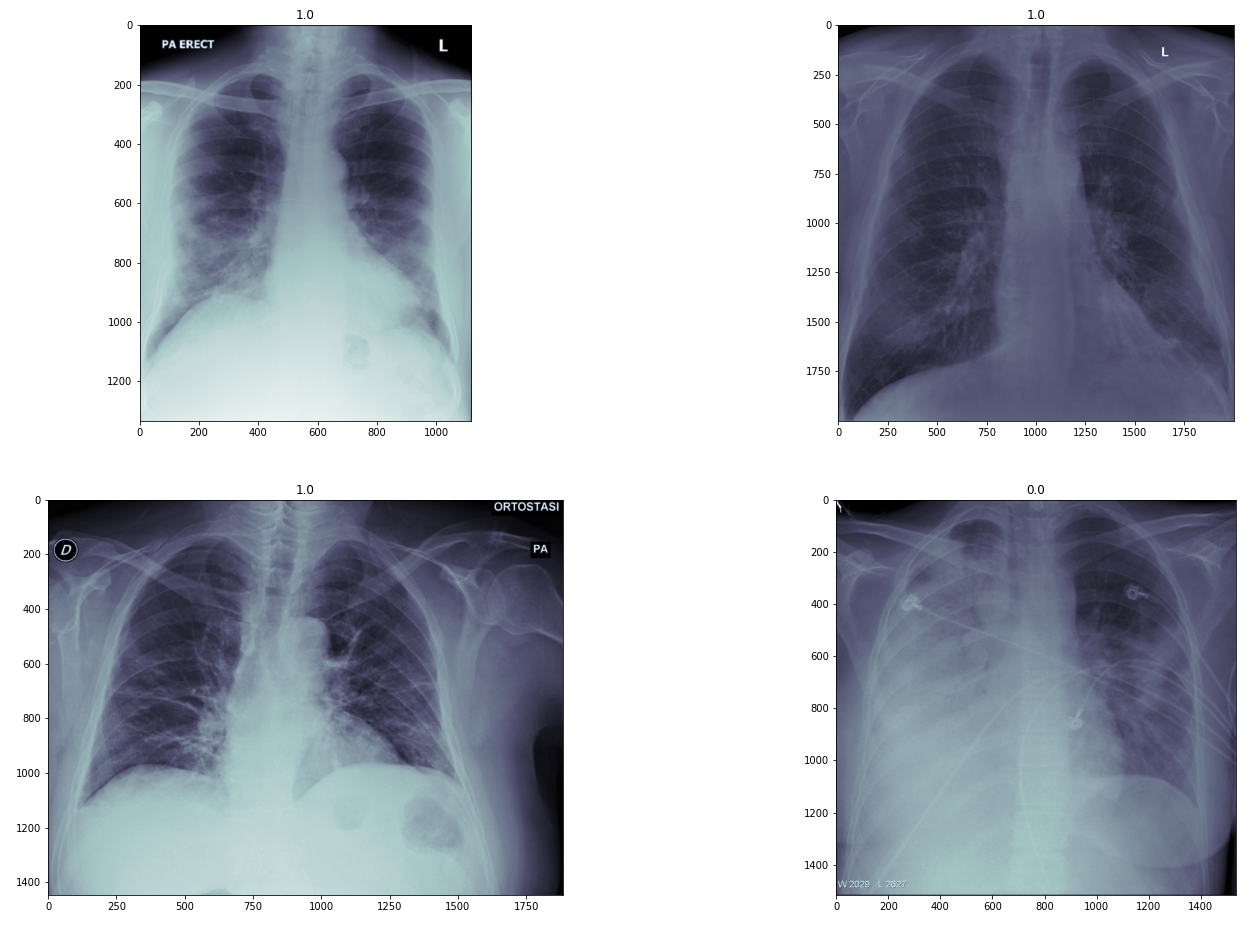}
\caption{\label{fig:xray-visual}1 means COVID-19 +Ve and 0 means -Ve.}
\end{figure}

We use covid-chestxray-dataset \cite{cohen2020covid} (Figure \ref{fig:xray-visual}), an open dataset collected from public and indirect collection from hospitals and physicians. The dataset is available on \href{https://github.com/ieee8023/covid-chestxray-dataset}{GitHub}.
It has a total of 535 AP and PA view of X-ray images in PNG format, which is a \href{https://guides.lib.umich.edu/c.php?g=282942&p=1885348#:~:text=PNG\%20(.&text=PNG\%20or\%20Portable\%20Network\%20Graphics,256\%20colors\%20supported\%20by\%20GIF.}{lossless image format}.
The ratio of COVID-19 positive and negative is 63.9\% and 36\%, where total positive labels are 342 and non-positives are 193. The metadata of this dataset contains labels of 25 lungs disease, shown in Table \ref{tab:pathalogy-data}.
To avoid issues with float round off the image pixels are \href{https://github.com/mlmed/torchxrayvision/issues/9}{normalized to be in range of [-1024,1024]}.

\begin{table}[h]
\centering
\begin{tabular}{l|c|r}
Pathalogy & Negative & Positive  \\\hline

Aspergillosis &	534 &	1\\
Aspiration & 534 &	1\\
Bacterial & 487 &	48\\
COVID-19 & 193 & 342\\
Chlamydophila	& 534& 1\\
Fungal	&512     &23\\
H1N1	&534&	1\\
Herpes 	&532&	3\\
Influenza	&531&	4\\
Klebsiella&	526&	9\\
Legionella	&526&	9\\
Lipoid	&527	&8\\
MERS-CoV&	527&	8\\
MRSA	&534	&1\\
Mycoplasma&	530&	5\\
No Finding	&520&	15\\
Nocardia	&531&	4\\
Pneumocystis	&513&	22\\
Pneumonia	&26	&509\\
SARS	&519&	16\\
Staphylococcus&	534&	1\\
Streptococcus	&518&	17\\
Tuberculosis	&524&	11\\
Varicella	&530	&5\\
Viral	&157	&378\\

\end{tabular}
\caption{\label{tab:pathalogy-data}covid-chestxray-datset pathology negative and positive frequency}
\end{table}

\section{Experiments}

\subsection{Diagnosing COVID-19 using Deep Representation Learning of CXRs}

Convolutional Neural Networks once trained on a huge dataset are frequently used for training on new task using transfer-learning \cite{transferlearn}, a method where weights of neural networks are initialised from the weights of pre-trained network. Deep Neural Networks learn the representation of the dataset and the data representation can be used for other tasks as well \cite{representationlearn}.
TorchXRayVision \cite{Cohen2020xrv}, an open-source library for Chest X-ray datasets and models is used for ease of experiment.
We use the DenseNet-121 \cite{huang2018densely} from \href{https://github.com/mlmed/torchxrayvision}{TorchXRayVision library} which is pre-trained on non-covid to produce 1024 dimensional vector representation of CXRs. 

The pre-trained DenseNet-121 model has been trained on large amount of chest x-ray dataset for different chest abnormality classification. We use this model for creating representation of COVID-19 CXR image data. We remove the last layer of the network, i.e. the classification layer, and extract the embedding from the last convolution layer. The obtained features are of dimension 7x7x1024, we applied 2D average pooling to convert it into 1x1x1024 dimension.
We use this embedding to train k-nearest neighbors classification algorithm (KNN) with scikit-learn Python library \cite{scikit-learn}, where $\boldsymbol{k}$ is the number of neighbors. We split the dataset into train and test using random sampling. Our train data consisted of 428 images and test data consisted of 107 images.

\begin{figure}[h]
\centering
\includegraphics[width=0.6\textwidth]{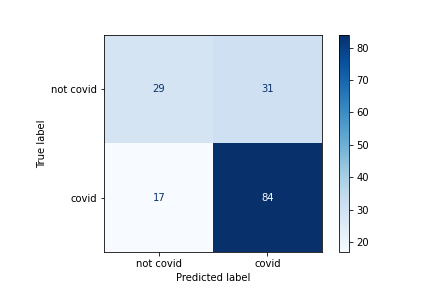}
\caption{\label{fig:knn-cmat}Confusion matrix of COVID-19 classification from CXR.}
\end{figure}

We train the KNN model with K=8 and took euclidean distance as distance metric $d = \sqrt{\sum(X_1 - X_2)^2}$.
We get 73\% precision and 83\% recall. We plot confusion matrix of our classification outputs in Figure \ref{fig:knn-cmat}.

\subsection{Predicting intubation requirement of patients}

\begin{figure}[h]
\centering
\includegraphics[width=0.6\textwidth]{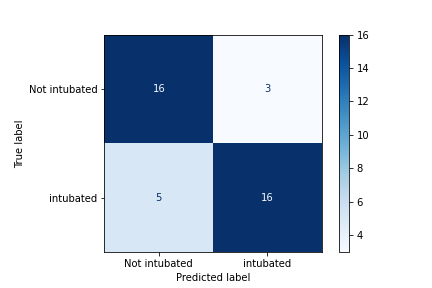}
\caption{\label{fig:intub-cmat}Confusion Matrix of patient intubation classification.}
\end{figure}

\cite{cohen2020covid} \textbf{intubated} attribute represents whether the patient was intubated at any point during his illness. Given a CXR, we predict whether the patient will need intubation, Figure \ref{fig:intub-cmat}.

For this task we used AP and PA view of Chest X-ray of COVID and other Chest diseases, total of 25 different pathology, Table \ref{tab:pathalogy-data}.
We produce our feature representation similar to section 4.1 and labels are 1 and 0 where 1 is intubation was required for the patient and 0 means intubation was not required.

\begin{figure}[h]
\centering
\includegraphics[width=0.6\textwidth]{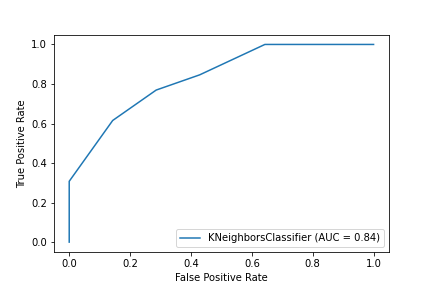}
\caption{\label{fig:intubation-auc-curve}Intubation classification ROC curve. 95\% Confidence interval for score 0.615, 0.863}
\end{figure}

We randomly split dataset into train and test set. Our training set comprised of 75\% of the total data. Train dataset length is 119 and test dataset length is 40.
We experiment with number of neighbors (k) from 2 to 14 inclusively. We find that k=5 gives the best precision and recall.
After training, we plot the confusion matrix of the test result as in Figure \ref{fig:intub-cmat}
20 intubated and 10 non-intubated patients are classified correctly as intubated and non-intubated respectively while 6 intubated and 4 non-intubated patients are classified incorrectly.
We calculate 0.84 (95\% CI 0.62, 0.86) AUC on the test set, Figure \ref{fig:intubation-auc-curve}.
We use bootstrap method to create 95\% CI with 10,000 bootstrap samples with replacement from the test dataset. We observe 0.84 (95\% CI 0.615, 0.863) ROC-AUC for our KNN classifier.
This shows how we can use representations when dataset is limited.

\section{Conclusion}
We need to leverage Deep learning methods to automate our healthcare system wherever possible. Deep learning methods can assist in better resource management.
X-rays are economical to CT scan and is available most of the places. It can be used to get an instant prognosis which can help in deciding quarantine and triaging.
Severity prediction can be used to check if the current resources can support the patient status. Intubation support prediction can help hospital in managing the bed, ventilator and oxygen cylinders in advance.

However, due to COVID-19 there has been a lot of research that has come up with different solution and suggestion for disease diagnosis. But before deploying any machine automated solution we must need to test it thoroughly for any kind of bias.

\bibliographystyle{apalike}
\bibliography{sample}

\end{document}